%% file: root.tex
\documentclass[letterpaper, 10 pt, conference]{ieeeconf}  
\pdfoutput=1

\IEEEoverridecommandlockouts                              

\usepackage{flushend}

\input{packages}
\input{macros}

\title{\LARGE \bf

Compensating the Absence of a Required Accompanying Person:  \\ A Draft of a Functional System Architecture for an  Automated  Vehicle*

	\thanks{\hspace{-1em}$^{*}$ This research was carried out as part of the UNICAR\emph{agil} project (FKZ 16EMO0285). We would like to thank the Federal Ministry of Education and Research (BMBF) for its financial support of the project and all members of the consortium for their contribution to this publication.}%
}

\author{Tobias Schräder$^{1}$,  Torben Stolte$^{1}$, Inga Jatzkowski$^{1}$, Robert Graubohm$^{1}$, Marcus Nolte$^{1}$ and  Markus Maurer$^{1}$
	\thanks{\hspace{-1em}$^{1}$Authors are with Institute of Control Engineering, TU Braunschweig, Germany
		{\tt\small \{schraeder, stolte, jatzkowski, graubohm,nolte, maurer\}@ifr.ing.tu-bs.de}}%
}
\begin{document}
	
	\onecolumn
		\Huge {IEEE copyright notice} \\ \\
	\large {\copyright\ 2021 IEEE. Personal use of this material is permitted. Permission from IEEE must be obtained for all other uses, in any current or future media, including reprinting/republishing this material for advertising or promotional purposes, creating new collective works, for resale or redistribution to servers or lists, or reuse of any copyrighted component of this work in other works.} \\ \\
	
	{\Large Published in \emph{2021 IEEE International Conference on Intelligent Transportation Systems}, Indianapolis, IN, USA, September 19--22, 2021.} \\ \\ 
	
	{\Large DOI: 10.1109/ITSC48978.2021.9564762} \\ \\ 

		\twocolumn

\maketitle

	\thanks{

\begin{abstract}

\input{sections/abstract}

\end{abstract}

\section{INTRODUCTION}

\input{sections/introduction}
\section{RELATED WORK}
\input{sections/funcarc}

\section{FUNCTIONAL SCOPE FOR THE AUTOMATION OF SUPPORTING ACTIVITIES}
\input{sections/requirements}
\section{DRAFT OF A FUNCTIONAL VEHICLE ARCHITECTURE}
\input{sections/draft}

\section{OPERATIONAL DESIGN DOMAIN}
\input{sections/outlook}
\section{OUTLOOK: THE AUTONOMOUS FAMILY VEHICLE ``AUTOELF"}
\input{sections/casestudy}

\section{CONCLUSION}
\input{sections/conclusion}

\renewcommand*{\bibfont}{\footnotesize} 

\printbibliography

\end{document}

%% file: packages.tex
\usepackage{graphicx}
\graphicspath{ {./figures/} }
\usepackage[table]{xcolor}
\usepackage{calc}
\usepackage{tikz}
\usetikzlibrary{shapes,
				automata,
				arrows,
				arrows.meta,
				matrix,
				backgrounds,
				fit,
				patterns,
				decorations.markings, 
				svg.path, 
				shapes.multipart, 
				shapes.geometric, 
				external, 
				shadows, 
				patterns,
				positioning, 
				calc, 
				decorations, 
				scopes,
				fadings,
				bending,
				scopes}
\usepackage{pgfplots}
\pgfplotsset{compat=1.17} 
\usepackage{pgfplotstable}
\usepackage{siunitx}
\usepackage[
			bibstyle=ieee,
			citestyle=numeric-comp,
			backend=biber,
			minnames=1,
			maxcitenames=2,
			maxbibnames=99,
			isbn=false,
			url=false,
			natbib=true,
			clearlang=true,
			date=year
			]{biblatex} 
\DeclareSourcemap{
	\maps[datatype=biber]{
		\map{
			\step[fieldsource=note, final]
			\step[fieldset=addendum, origfieldval, final]
			\step[fieldset=note, null]
		}
	}
}
\usepackage{xpatch}
\xpatchbibdriver{online}
{\printtext[parens]{\usebibmacro{date}}}
{\iffieldundef{year}
	{}
	{\printtext[parens]{\usebibmacro{date}}}}
{}
{\typeout{There was an error patching biblatex-ieee (specifically, ieee.bbx's @online driver)}}

\usepackage{lipsum}
\usepackage{amsmath}
\usepackage{amssymb} 	

\usepackage{amsthm} 	

\newtheoremstyle{defn}
{3pt}
{3pt}
{}
{\parindent}
{\bfseries}
{.}
{.5em}
{}

\theoremstyle{defn}

\usepackage{bm} 		
\usepackage{booktabs} 	
\usepackage{csquotes}
\usepackage{microtype}
\usepackage{fmtcount}
\usepackage[hidelinks]{hyperref}



%% file: macros.tex
\makeatletter
\newcounter{IEEE@bibentries}
\renewcommand\IEEEtriggeratref[1]{%
	\renewbibmacro{finentry}{%
		\stepcounter{IEEE@bibentries}%
		\ifthenelse{\equal{\value{IEEE@bibentries}}{#1}}
		{\finentry\@IEEEtriggercmd}
		{\finentry}%
	}%
}
\makeatother

\ifx \isaccepted \undefined
\newcommand\copyrighttext{%
	\footnotesize \centering This work has been submitted to the IEEE for possible publication.\\ Copyright may be transferred without notice, after which this version may no longer be accessible.}
\else
\newcommand\copyrighttext{%
	\footnotesize \parbox[t]{.11\textwidth}{\copyright{} \the\year~IEEE.} \parbox[t]{.89\textwidth}{Personal use of this material is permitted. Permission from IEEE must be obtained for all other uses, in any current or future media, including reprinting/republishing this material for advertising or promotional purposes, creating new collective works, for resale or redistribution to servers or lists, or reuse of any copyrighted component of this work in other works.}}
\fi
\newcommand\copyrightnotice{%
	\ifx \compileforpublish \undefined
	\else
	\begin{tikzpicture}[remember picture,overlay]
	\node[anchor=south,yshift=10.5pt] at (current page.south) {\parbox{\dimexpr\textwidth-\fboxsep-\fboxrule\relax}{\copyrighttext}};
	\end{tikzpicture}%
	\fi
}

%% file: sections/abstract.tex
A major challenge in the development of a fully automated vehicle is to enable   a large variety of users to use the vehicle independently and safely. Particular demands arise from user groups who rely on human assistance when using conventional cars. For the independent use of a vehicle by such groups, the vehicle must compensate for the absence of an accompanying person, whose actions and decisions ensure the accompanied person's safety even in unknown situations. The resulting requirements cannot  be fulfilled only by the geometric design of the vehicle and the nature of its control elements. Special user needs must be taken into account in the entire automation of the vehicle. In this paper, we describe  requirements for compensating the absence of an accompanying person and show how  required functions can be located in a hierarchical functional system architecture of an automated vehicle.  In addition, we outline the relevance of the vehicle's operational design domain in this context and present a use case for the described functionalities.

%% file: sections/introduction.tex
\label{sec:introduction}
Automated vehicles promise to provide more autonomy for people who are not able to operate a conventional car by themselves due to physical or mental restrictions \cite{harperEstimatingPotentialIncreases2016a,tremouletTransportingChildrenAutonomous2020}. For individual transport, these people are dependend on the availability of a companion who is fit to drive. Depending on the needs of the accompanied person, the use of a conventional car may require supporting activities of an accompanying person that go beyond the actual driving task. An accompanying person might provide physical support to people with motor impairments to ensure a safe entry and exit from a car. Moreover, activities of an accompanying person might also be needed when children, who are unable to move safely in road traffic on their own, get in and out of the car. Furthermore, it can be assumed that an accompanying person recognizes hazards, makes decisions  in the interest of the accompanied persons, and takes actions that ensure the safety of their passengers. 
For those who  dependent on human assistance when using a conventional car, it would be a significant advantage if they were able to use an automated vehicle independently without facing an increased risk. For this purpose, an automated vehicle must compensate for the absence of an accompanying person.

In the past, concepts for fully automated vehicles have been presented that include approaches for the use by persons who are unable to drive conventional vehicles. For example, IBM has presented a vehicle that can be entered by elderly persons with motor impairments via a ramp and operated with the aid of a speech computer \cite{ibmAccessibileTransportationIBM2018}. Moreover, Waymo has already demonstrated how blind people can use  automated road vehicles independently~\cite{halseyBlindManSets2016}.

However, the known approaches for the use of driverless vehicles by  persons with limited abilities are essentially limited to the geometry and design of the human-machine interface of the respective vehicles. A vehicle automation that allows a highly diverse group of passengers to travel as safely and  efficiently as with an accompanying person, even in critical situations, has not been presented so far. We assume that an automated vehicle is supposed to behave as if it were operated by a caregiving person for the safe transportation of vulnerable passengers. Therefore, a comprehensive consideration is needed in the design of the entire vehicle automation system. Nevertheless, to date the focus in the design and description of driverless vehicle's system architectures has been primarily on the automation of the driving task. Accordingly, the complex  activities of a car driver in his or her capacity as an accompanying person have  been disregarded in system architectures for automated vehicles so far.

In this paper, we first give a brief overview of  previously published functional architectures for driverless vehicles and other robotic systems with focus on the included consideration of users. In Section~\ref{sec:requirements}, we outline functional requirements for a driverless vehicle that compensates for the absence of an  accompanying person during vehicle use. In Section~\ref{sec:draft}, we describe the draft of a functional vehicle architecture that is structurally based on existing approaches and contains an extended functional scope. 
Afterwards, we outline the extension of the  vehicle's operational design domain in Section~\ref{sec:outlook}. In Section \ref{sec:casestudy}, we present a concrete use case of an automated vehicle that requires functionalities considered in this paper and allows us to validate our concepts.

%% file: sections/funcarc.tex
\label{sec:funcarc}
The description of system architectures is defined in the ISO/IEC/IEEE 42010 standard \cite{ISOIECIEEE}. An explicit description of a system's architecture can be an important tool for the development of more complex systems. Fundamental concepts of a technical system can be represented by architecture descriptions \cite{ISOIECIEEE}. Thereby, different views of a system's architecture can be distinguished, such as a hardware view, a software view, a capability view, or  a functional view  \cite{bagschikSystemPerspectiveArchitecture2018}.  Functional system architectures display the functional units of a system and their interfaces to each other and to external systems. They are independent of concrete components and can form a basis for the new development of a complex system. In this section, we provide an overview of the contents of functional system architectures  for automated road vehicles and the contained considerations of passenger needs. In addition, we give a brief overview of other robotic systems and their functional architectures that contain approaches that are relevant for our work.

Numerous functional system architectures for automatically driving vehicles have been presented, for example in~\cite{dickmannsSeeingPassengerCar1994,maurerFlexibleAutomatisierungStrassenfahrzeugen2000,bayouthHybridHumanComputerAutonomous1998,tasFunctionalSystemArchitectures2016,lotzReferenzarchitekturFurAssistierte2017,matthaeiAutonomousDrivingTopdownapproach2015,ulbrichFunctionalSystemArchitecture2017,kunzAutonomousDrivingUlm2015,behereFunctionalArchitectureAutonomous2015,joDevelopmentAutonomousCar2014,torngrenm.ArchitectingSafetySupervisors2018}. They differ in their structure, their level of detail, and their interfaces to external systems. One common feature of all functional architectures for automated vehicles known to us are explicitly described modules for perceiving the vehicle's environment.  The information used for perceiving the environment  can be derived from processed data of the vehicle's own sensors, but also from external sources such as a communication interface to the infrastructure. In addition, some of the  functional  architectures provide modules for the self-perception of a vehicle \cite{matthaeiAutonomousDrivingTopdownapproach2015,ulbrichFunctionalSystemArchitecture2017, lotzReferenzarchitekturFurAssistierte2017} or for monitoring the performance of individual modules \cite{tasFunctionalSystemArchitectures2016}.

Depending on the concrete architecture, the information available to the vehicle is combined and processed in  a more or less abstracted way. In some   functional architectures for automated vehicles, all information is represented in a single world model~\cite{lotzReferenzarchitekturFurAssistierte2017,matthaeiAutonomousDrivingTopdownapproach2015,ulbrichFunctionalSystemArchitecture2017,kunzAutonomousDrivingUlm2015,behereFunctionalArchitectureAutonomous2015}. 
The  collection of data from which information about a passenger can be obtained, the perception of a passenger’s state or representation of passengers is not considered in the descriptions of the
known functional system architectures that are developed for automating the driving task.  One exception is the concept presented in \cite{drewitzUserfocusedVehicleAutomation2020} for taking  passengers’ needs into account in order to reduce their subjective uncertainty by adapting several vehicle functions.

Functional modules for the planning and the execution of the driving task differ in their structure and the way of their description as well. Some of the architectures~\cite{lotzReferenzarchitekturFurAssistierte2017,matthaeiAutonomousDrivingTopdownapproach2015,ulbrichFunctionalSystemArchitecture2017} are inspired by the hierarchical structure of a model of the human driving task~\cite{dongesConceptualFrameworkActive1999}. In some modules for mission execution, interfaces to vehicle users are described. For example, the functional system architecture described in \cite{maurerFlexibleAutomatisierungStrassenfahrzeugen2000} provides an interface for a mission input and the system architectures described in~\cite{bayouthHybridHumanComputerAutonomous1998,lotzReferenzarchitekturFurAssistierte2017,matthaeiAutonomousDrivingTopdownapproach2015,ulbrichFunctionalSystemArchitecture2017} include interfaces for the input and output of information at different hierarchical levels of the mission execution. Furthermore, the architecture described in~\cite{joDevelopmentAutonomousCar2014} provides an interface for developers.

Some functional descriptions of system architectures for automated vehicles  include modules for controlling the actuation system  which is composed of a steering system, a drivetrain and a braking system \cite{lotzReferenzarchitekturFurAssistierte2017,matthaeiAutonomousDrivingTopdownapproach2015,ulbrichDefiningSubstantiatingTerms2015}. However, other actuators of a vehicle that are not directly required for driving the vehicle -- such as a door locking system -- are  not considered in  detail in this context. A model of driver tasks presented by \textcite{bubbFutureApplicationsDHM2007} includes tertiary tasks of a car drivers that are not needed for driving, but serve the purposes of comfort, communication, information, and entertainment.  However, even this model does not describe the activities of  drivers in their capacity as  accompanying persons.

Apart from automatically driving vehicles, numerous  architectures of technical systems have been developed in the past decades in which interaction with the user and the consideration of user needs are of crucial importance. For instance, user interaction in the  system architectures of social robots is of central importance.
In \cite{bartneckDesigncentredFrameworkSocial2004}, a social robot is defined as a robot that acts autonomously, interacts with humans -- typically in a cooperative way -- and follows social norms. These functionalities require the robot to perceive its environment,  its users, to interpret situations, to plan actions, and to execute the planned actions. Some of the social robots are designed to appear anthropomorphic and can communicate multimodally~\cite{bartneckDesigncentredFrameworkSocial2004}. In~\cite{malfazNewArchitectureAutonomous2004}, a system architecture is described that allows a robot to communicate its state in  form of emotions.  In this context, a social robot can be understood not only as a tool to be controlled, but also as a "counterpart"~\cite{gorostizaMultimodalHumanRobotInteraction2006}.
The use cases of social robots are diverse, for example in entertainment,~\cite{aaltonenHelloPepperMay2017}, for productivity enhancement~\cite{lenzMechanismsCapabilitiesHuman2014}, or in caregiving~\cite{broekensAssistiveSocialRobots2009}.


In the functional system architectures of social robots, models of  human cognition  are often used as an archetype. 
The architectures described in~\cite{barberNewHumanBased2001, malfazBiologicallyInspiredArchitecture2011, lazzeriDesigningMindSocial2018}, for example, include a division of functional units into a planning and a reactive system level. Simple, less computationally intensive  skills of the robot are located at a lower level, hardware related system level, whereas the more complex skills are located at a higher level.  In a similar way, the architecture described in~\cite{bonassoExperiencesArchitectureIntelligent1997} is divided into three levels, with a planning level as a top level and a bottom level that sends control commands to the robot's actuators.

All in all, the presented system architectures for social robots take specific users' needs more into account than the known functional system architectures  for automated vehicles. The focus of the known  functional system architectures for automated vehicles has so far been on the automation of the driving task. Vehicle users have typically been considered only as recipients of information or operators of the vehicle, but not as vulnerable passengers who depend on the vehicle's support and protection. 
Parallels between social robots and automated vehicles can be seen in the hierarchization of functional architectures when these are based on human activities.
In addition, the  system architectures for social robots show far-reaching approaches for the communication between humans and an autonomous machine. The vehicle considered in this paper is intended to be such an autonomous machine, interacting with the users and reacting to them if necessary in order to compensate for the absence of an accompanying person.

%% file: sections/requirements.tex
\label{sec:requirements}
In a first approach to derive  functional requirements for a vehicle that compensates for the absence of a previously required accompanying person, we focused on the activities of an accompanying person when using a conventional car~\cite{schraderApproachRequirementAnalysis2019}.  
These required activities are determined by the limitations of the vehicle user's abilities, which are their personal prerequisites for independently accomplishing a task \cite{wirtzDorschLexikonPsychologie2014}. In some cases, it is possible that the decisions and actions of an accompanying person do not match the intentions of the accompanied person. 

Accompanying a person in need of support can have an impact on the primary, secondary and tertiary tasks  of a car driver, which are described in~\cite{bubbFutureApplicationsDHM2007}. Further requirements arise when transporting persons with special protection needs in  public. For instance, \textcite{tremouletTransportingChildrenAutonomous2020} as well as \textcite{leeParentsPerspectivesUsing2018} mention parents' expectations of an automated vehicle so that they would let their children travel alone in the vehicle.

Some requirements that result from the compensation of an accompanying person's absence can be fulfilled by  existing solutions in the design of a vehicle. For example, the operability of the vehicle can be ensured by the design of the vehicle's controls even if the perceptive and sensomotoric abilities of the intended users are limited. Approaches for designing and structuring the human-machine interface of an automated vehicle have been previously presented~\cite{benglerHMIHMIsHMI2020}.
Furthermore, supporting activities that are required when accompanying a  locomotor-impaired person can be at least partially compensated for by the geometric design of the vehicle and additional auxiliary devices, such as an automated loading aid.

A major challenge, however, is to compensate for the loss of perceptive and cognitive activities of an accompanying person. This means that the vehicle, like an accompanying person, must have the ability to recognize and react to the needs of its passengers and, above all, to hazards for the passengers.  In the case of an accompanying person, it can be assumed that they will  react to unknown disturbances  in a way that harm to passengers will be prevented  or at least keep it to a minimum. An automated vehicle that can be used by people who have so far been dependent on human assistance when using a conventional car must therefore be at least as robust as a human companion in the face of  disturbing influences.

\begin{figure}[h]
	\centering
	\includegraphics{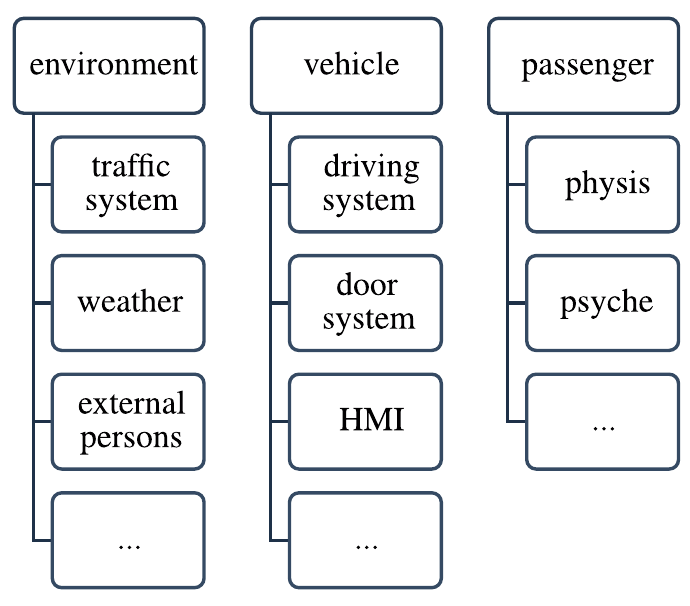}
	\caption{Categorization of disturbances}
	\label{fig:disturbances}
\end{figure}

Disturbances in the operation of the vehicle can arise from outside the vehicle, i.e. from the vehicle's environment, from the vehicle itself, or from the passengers. Fig.~\ref{fig:disturbances} shows a categorization of possible causes of disturbances at the two highest levels of abstraction.  Further specification -- especially in the difficult to delimit categories "environment" and "passenger" -- leads to an unlimited number of possible disturbances. In addition, all disturbances can  occur in combination, which  increases the complexity of the system's development.  Moreover, it is possible that some disturbances only cause hazards during independent vehicle use in combination with certain user characteristics. For example, a malfunctioning door actuator would not lead to passengers being  trapped in the vehicle if a manual emergency redundancy was provided. However, if the vehicle's passengers are unable to operate the manual emergency actuation due to physical limitations, this redundancy is eliminated and users cannot escape from the vehicle in case of a malfunctioning door actuator.

To compensate for the absence of an accompanying person, the automated vehicle's perception system must be able to perceive the needs of passengers and potential hazards to them. The vehicle must put itself, its environment and the user in context and represent this contextual information.  At the same time, the mission execution of the vehicle's automation system must have the ability to take recognized needs and hazards into account when a mission is planned and executed. Thereby, the vehicle must be functional in an unlimited number of scenarios, just as in the automation of the driving function without incorporation of specific user's needs.

%% file: sections/draft.tex
\label{sec:draft}

\begin{figure*}[h]
	\centering
	\scalebox{1}{\includegraphics{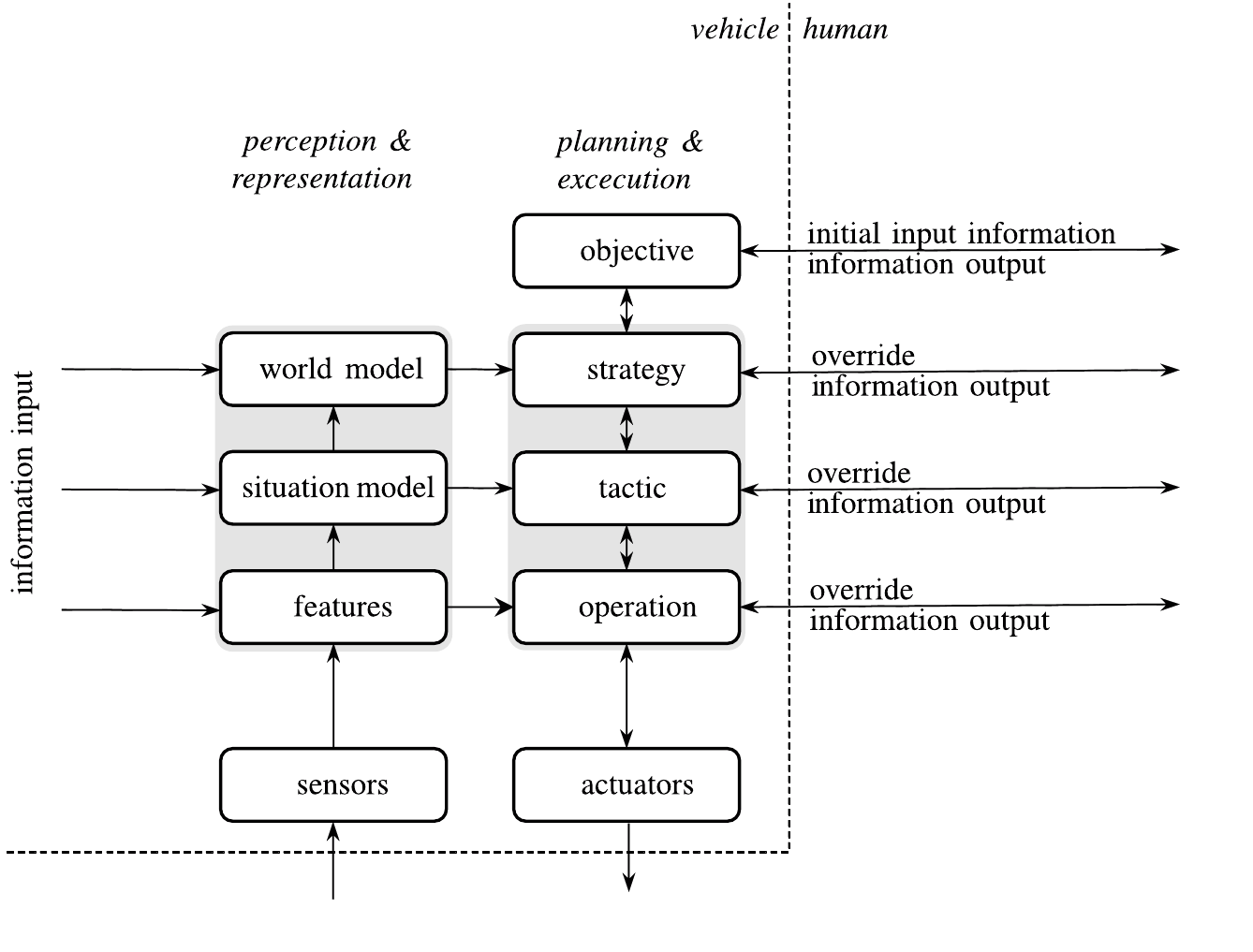}}
	\caption{Draft of a functional vehicle architecture for compensating for the absence of an accompanying person,  based on \cite{matthaeiAutonomousDrivingTopdownapproach2015}}
	\label{fig:architecture}	
\end{figure*}

As some of the already known system architectures for automating the driving task \cite{lotzReferenzarchitekturFurAssistierte2017,matthaeiAutonomousDrivingTopdownapproach2015,ulbrichFunctionalSystemArchitecture2017} and the presented social robots, the draft of a system architecture presented in this paper is functionally divided into a subsystem for perceiving and representing information and a subsystem for mission planning and  executing. Both subsystems are horizontally arranged side by side and connected via several interfaces as shown in Fig.~\ref{fig:architecture}.

To cope with the diversity and complexity of tasks, a hierarchical decomposition of a task into subtasks is commonly applied in robotics \cite{dillmannHierarchischeModellierungRobotersteuerungsarchitekturen1991}. As mentioned above, some system architectures of social robots that use humans as models contain such hierarchical structures \cite{gorostizaMultimodalHumanRobotInteraction2006,barberNewHumanBased2001,malfazBiologicallyInspiredArchitecture2011}. 
Some of the system architectures that are developed for the automation of the driving task presented so far  use a hierarchical structure as well. The architectures described in \cite{lotzReferenzarchitekturFurAssistierte2017, matthaeiAutonomousDrivingTopdownapproach2015,ulbrichFunctionalSystemArchitecture2017} are based on the hierarchical decomposition of the human driving task according to \textcite{dongesConceptualFrameworkActive1999}.  Thereby, a distinction is made between a strategic, a tactical, and an operational level. A significant difference between the individual levels are the runtimes of the subtasks located within these levels. In the same way, decisions and actions of an accompanying person who drives a car can be divided into three levels and assigned to  time horizons of different lengths. Accordingly, the draft of a functional system architecture described in this paper provides a vertical division of both subsystems  into three hierarchical levels, which are described in more detail hereafter.

\subsection{Mission Objective}
The initial input into the system is the mission objective, which is specified by the vehicle itself or by a human operator, as described in existing approaches \cite{maurerFlexibleAutomatisierungStrassenfahrzeugen2000,matthaeiAutonomousDrivingTopdownapproach2015,ulbrichFunctionalSystemArchitecture2017}. If the mission objective is set by a human operator, it can be either a passenger or an external person, depending on the use case of the vehicle. 
A mission objective can contain more information than just locations to be reached: For example, information about the passengers to be transported or other constraints -- which can be used to distinguish between a particularly urgent trip and a leisure trip -- may also be required for the execution of a mission by the automated vehicle.

\subsection{Strategic System Level}
At the strategic level, a long-term plan for reaching the mission objective is created. In accordance with the hierarchical division of the human driving task as described in~\cite{dongesConceptualFrameworkActive1999}, the route to a selected destination is planned. If the functional scope of the vehicle's automation has to compensate for the absence of an accompanying person, further tasks with longer time horizons can be assigned to the strategic system level. This includes considering user-individual needs when selecting a route and  deciding if a mission's objective needs to be changed. For example, the potential disrupting influences to vehicle's operation indicated in Fig.~\ref{fig:disturbances} may result in an original mission objective no longer being achievable without harming a passenger. In such cases, the top level of mission execution must be able to develop a plan that minimizes harm to a passenger. Such a plan may include, for example, returning to the departure point, driving to a different destination than initially planned, stopping immediately, or alerting external parties such as a control room or rescue service.  Therefore, the automated vehicle must  be able to predict the longer-term effects of possible courses of action on its passengers, which may extend beyond the duration of the mission.

The strategy to reach the mission objective is based on  the initial mission objective, the performance reported by the hierarchically subordinate module for tactical mission execution, and the analysis and prediction of a world model. This world model contains an abstract representation of the vehicle, its environment, and the passengers, which are spatially and temporally related to each other. All available information about the environment, the vehicle, and the passengers that is relevant for the entire mission execution in the long term is represented in this model. This includes, for instance, map data containing passenger-related information on locations and areas, the vehicle's current range, the expected driving time, and spatial distances to certain locations, or the expected long term passenger status, which in turn can correlate with the maximum possible duration of a passenger's stay in the vehicle.

\subsection{Tactical System Level}
The plan created by the strategic mission execution module is executed by the hierarchically subordinate tactical mission execution module. 
 The hierarchical division of the driving task according to \textcite{dongesConceptualFrameworkActive1999} foresees the decision on driving maneuvers and the planning of the vehicle's movement at the tactical level. In the system considered here, passenger needs beyond kinetosis prevention, driving enjoyment, and comfort must be considered when planning the vehicle's movement. For example, special requirements of motor impaired passengers for an entry or exit point are possible in this context. In addition, the timing for unlocking or automatically opening and closing the vehicle's doors is to be determined at the tactical system level. Thereby, the unlocking and automated opening of a vehicle's door can be further delayed due to current parallel traffic. At the same time, the tactical mission execution module reports its own performance to the hierarchically superior module. This means, for instance, that a decision must be made at the strategic level about  changing the destination or other available options when the vehicle's doors cannot be opened due to traffic or no suitable stopping location is available at the destination.

To implement the plan provided by the strategic mission execution module, the tactical mission execution module accesses, analyzes, and predicts a detailed model of the current situation. According to the definition given by \textcite{ulbrichDefiningSubstantiatingTerms2015} in the context of the automation of a vehicle's driving function, a situation includes dynamic elements, scenery, a self-representation, and a representation of actors and observers.  Actors can be passengers who are put in a context to the vehicle and the environment of the vehicle.  A concept for the representation of passengers in a situation model is described in \cite{drewitzUserfocusedVehicleAutomation2020}, for example. When modeling the vehicle's environment, it is also possible to represent persons who are assigned a role different from that of other road users depending on their relationship to the vehicle's passengers.  For example, external persons may be requested to perform supporting activities or, in turn, may be a potential disruptive influence on the mission execution.

\subsection{Operational System Level}

The input provided by the tactical mission execution module is converted into control commands for the vehicle's actuators by the underlying operational mission execution module. The basis for this are the current performance reported by the actuator system and the features provided by the perception system. The vehicle's perception system filters features of the environment, the vehicle itself, and the passengers from the data available to the vehicle. Features relating to the passengers can be extensive -- even if only a few sensors are used. Contactless perception using cameras and microphones can  detect speech, identities, gestures and body postures \cite{mitraGestureRecognitionSurvey2007}, body temperatures \cite{gadeThermalCamerasApplications2014}, heart rates~\cite{leeContactFreeHeartRate2012}, breathing rates, e.g. cf.~\cite{bartulaCamerabasedSystemContactless2013}, or that a passenger has fallen~\cite{vishwakarmaAutomaticDetectionHuman2007}. In addition to features obtained from processed sensor data, features can be supplied to the system via an interface. This can be, for example, pre-stored information about passengers, such as special needs for vehicle use, or information that is transmitted to the vehicle from external persons, other vehicles, a control room, or an  infrastructural entity, such as a traffic control system.

\subsection{Sensors and Actuators}
To illustrate the scope of the system as well as the perception and action options available to the vehicle's automation system, the vehicle's sensors and actuators as depicted below the operational system level in Fig.~\ref{fig:architecture} are described  in the following.

Some of the sensors are used to perceive the vehicle's environment and are essential for automating the driving function of a vehicle. Another part of the sensor system is used to detect the vehicle's state. The vehicle state includes not only the driving state of the vehicle but also the states of the vehicle's  subsystems. This  includes even those subsystems that do not merely fulfill the driving function. These can be, for example, the door system or other subsystems located in the vehicle's interior. Another part of the vehicle's sensors is used to detect the state of the passengers. The passengers' state can include, for instance, indicators for their state of health or their position and movement within or in the vicinity of the vehicle. In addition, some of the controls -- by which passengers can communicate a concern to the vehicle and the vehicle can receive information about the passenger's condition that needs to be interpreted -- can also be included in this group of sensors. In addition, each sensor can fulfill multiple tasks.  For example, a thermometer not only provides information about the temperature of the vehicle's interior or the vehicle's environment, but also about the ambient temperature of a passenger, which in turn, allows conclusions to be drawn about the passenger's condition in combination with other information about them. 

The actuation system of an automated road vehicle includes at least a drivetrain, braking system, and steering system  required for the driving function. The actuation system of a vehicle assumed here also includes components that are not directly required for the driving function. One example of such components is the door system and its locking system. It enables passengers to exit or enter the vehicle and can prevent access by strangers. In addition, depending on the vehicle's equipment, other systems such as active entry or loading aids are possible as part of the vehicle's actuation system. The actuation system not only affects the state of the vehicle and thus, if necessary, the state of the vehicle's environment, but also the vehicle's passengers.

\subsection{Human Vehicle Interfaces}
The draft of a system architecture considered here provides several interfaces for passengers and external persons.  In the vehicle architectures described in~\cite{lotzReferenzarchitekturFurAssistierte2017,matthaeiAutonomousDrivingTopdownapproach2015,ulbrichFunctionalSystemArchitecture2017}, the subsystem for mission execution has interfaces for information output at all hierarchical levels. The type of information output is related to the type of tasks performed within each module. For example, at the strategic level, the vehicle's long-term plan can be made transparent to passengers.

In particular, if the mission objective is changed by the vehicle's automation, this information should be communicated to the users in order to maintain their trust in the vehicle. At the same time, it should be considered that the long-term, strategic planning of the vehicle can also be relevant for users who are currently outside the vehicle, while the tactical planning must also be partially visible to other road users.
Moreover, concrete information or simple warnings can be provided at the operational system level.

Furthermore, the input and override options at the respective hierarchical levels of the vehicle differ according to the time horizons of the individual levels. The actual use of these interfaces depends, for example, on the use case of the automated vehicle or its operating modes. It is conceivable that passengers could be  allowed to intervene only at certain hierarchical levels of the system, that they or external persons are only allowed to intervene under certain conditions, or that persons can even be requested to provide control interventions.  Under certain circumstances, it may also be necessary for the vehicle to make the contents of its representation models available to a control room  or for the realization of parental control features as described in~\cite{tremouletTransportingChildrenAutonomous2020}.

 As in the case of the information output, the nature and design of the override options can be highly dependent on the particular vehicle. Conceivable approaches are those described in \cite{benglerHMIHMIsHMI2020}, or even the integration of a social robot, as described in \cite{gorostizaMultimodalHumanRobotInteraction2006}, for communication with the passenger  at all hierarchical levels of the system described above. For that purpose, further functional modules need to be added to the right side of the architecture shown in Fig.~\ref{fig:architecture}. Depending on the needs of the passengers, the design of the interface can be a decisive criterion for the operability and thus the independent usability of the vehicle.  However, the further specification of the user interface goes beyond the description of a functional system architecture in this paper.

%% file: sections/outlook.tex
\label{sec:outlook}

For the realization and safe operation of an automated vehicle, the definition of the intended operational design domain can be of  importance~\cite{koopmanHowManyOperational2019}. 
Common restrictions to date include a possible use of an automated vehicle only on highways, exclusively in urban areas, or only in certain speed ranges.
For the more far-reaching automation of a vehicle as described in this paper, additional dimensions of the defined operational domain are conceivable. For instance, requirements for the persons potentially traveling alone in the vehicle are necessary. In addition to  formal requirements, such as a minimum age, there could also be requirements for a person's physical and cognitive abilities. A set of requirements for children to use an automated vehicle on their own  mentioned by a group of parents is described by~\textcite{tremouletTransportingChildrenAutonomous2020}. It can be assumed that there are  characteristics that, from today's perspective, make it impossible for  affected persons to remain in an automated vehicle on their own under all circumstances. However, the extent to which assumptions about users can be formalized remains to be clarified.

Moreover, there may also be passenger-dependent restrictions with regard to the spatial area in which the vehicle is used. It is conceivable, for example, that the vehicle may only move in the areas known to the passenger on board. It is also conceivable that areas may not be considered as possible locations for the vehicle due to increased crime rates. In addition, the distance or the maximum possible driving time can be restricted depending on the intended passengers. Likewise, the vehicle's  operational design domain may be linked to certain environmental conditions or the availability of a potential human assistant within a defined distance.

Another aspect when defining the  operational design domain  is the vehicle's specific use case. The use of a vehicle within a known user group -- as in the case of the autonomous family vehicle introduced in Section~\ref{sec:casestudy} -- provides certain advantages: important information about individual users can be stored in advance in a user profile, the potential administrator of the vehicle knows the passengers, a familiarity of the users with the vehicle can be assumed, some users within the user group can still perform parts of their tasks as a former companion, and vandalism or a threat by criminals within the vehicle -- unlike with public transport -- can be mostly excluded.

%% file: sections/casestudy.tex
\label{sec:casestudy}
As one of four use cases of driverless vehicles, an autonomous family vehicle is being designed and prototyped over the course of the project UNICAR\emph{agil} \cite{woo18}. One objective within the project is to develop a vehicle that takes over assisting tasks of a previously required accompanying person. Thus, it becomes an autonomous family assistant, as indicated by the vehicle's name \emph{auto}ELF. The aim of independent use  by as many members of a multi-generational family as possible results in novel requirements for the automation of the vehicle. By assuming a concrete use case, the requirements can be specified in more detail and approaches for the use-case-specific challenges can be elaborated. At the same time, numerous challenges are uncovered through the prototypical realization, for example in the inclusion of users who are dependent on a walking aid or conflicts of interest between different stakeholder groups due to novel functionalities of the vehicle~\cite{graubohmValueSensitiveDesign2020}.

The vehicle, which is primarily developed by universities, will be equipped with an automated door system, a barrier-free entry, interchangeable control elements, a flexibly usable and ground-level accessible storage space for wheelchairs, strollers, walking aids, and other items that are difficult to store in a conventional vehicle. Due to the dimensions of the vehicle's body and the interchangeability of interior trim elements, this experimental vehicle offers  the opportunity to integrate a wide variety of sensors and further appliances into the interior. 
The vehicle's hardware thus provides the prerequisites for independent use by persons who were previously supported by an accompanying person.  Thereby, the vehicle can be the basis for implementing the described automation system at all three hierarchical system levels. 

 An illustrative example for the allocation of a accompanying person's tasks in the presented draft of a system architecture can be seen when persons who cannot climb a step on their own are considered as independent users: the experimental vehicle's actuation system includes an actuated platform that   can lift the passenger over the vehicle's sill. The movement of the platform, which is equipped with several sensors, and the automated  door system are  controlled by the operational system level. At the tactical system level, the operation of the platform is triggered and the vehicle is maneuvered to a position suitable for using the platform. At the system's top level, locations that are generally unsuitable for the use of the platform due to their topography are excluded as possible stopping points. In addition, the increased boarding and deboarding time when transporting a passenger who relies on the platform is taken into account when planning the vehicle's missions with the help of a family app.

Another example that illustrates  assignment of accompanying person's task in a hierarchical system architecture is  the concept for \emph{auto}ELF's reaction on a detected medical emergency of a passenger. At the strategic system level, a decision is made  whether and to which destination the vehicle should continue to drive, whereas at the tactical system level, the vehicle's behavior is adapted to the current situation -- which also includes the passenger's current attributes known to the vehicle.

%% file: sections/conclusion.tex
By compensating an accompanying person's supporting activities, the independent use of a driverless vehicle can be provided to new groups of users. Taking on these complex tasks can be a significant challenge for the entire automation system of a driverless vehicle. A look at previously published functional architectures of driverless vehicles shows that their development primarily pursued the automation of the driving task, but not the supporting tasks  of an accompanying person beyond the actual driving task.

In this paper, we describe how tasks of an accompanying person that have not been considered so far can be placed in a draft of a functional system architecture for an automated vehicle.  Thereby, complex tasks are hierarchically structured and a first, albeit still  abstract, functional outline for the realization of an experimental vehicle is provided. A hierarchical division of the system  -- as it was previously described in functional system architectures of driverless vehicles -- appears to be appropriate to handle the complexity of the  considered tasks. At the same time, however, it also becomes clear how comprehensive and complex the vehicle's functions required to compensate for the loss of an accompanying person actually are and what challenges still have to be overcome in their implementation. The extent of these challenges  depends on the assumed operating conditions of the vehicle, which  include the characteristics of the potential passengers.